\documentstyle[12pt]{article}

\def\hybrid{\topmargin -20pt  \oddsidemargin 0pt
      \headheight 0pt   \headsep 0pt
      \textwidth 6.25in 
      \textheight 9.5in 
      \marginparwidth .875in
      \parskip 5pt plus 1pt   \jot = 1.5ex}

\hybrid

\begin{document}
\def\x{\times}
\def\beq{\begin{equation}}
\def\eeq{\end{equation}}
\def\beqa{\begin{eqnarray}}
\def\eeqa{\end{eqnarray}}

\sloppy
\newcommand{\be}{\begin{equation}}
\newcommand{\eq}{\end{equation}}
\newcommand{\ov}{\overline}
\newcommand{\un}{\underline}
\newcommand{\p}{\partial}
\newcommand{\la}{\langle}
\newcommand{\ra}{\rangle}
\newcommand{\bl}{\boldmath}
\newcommand{\ds}{\displaystyle}
\newcommand{\nl}{\newline}
\newcommand{\th}{\theta}



\renewcommand{\thesection}{\arabic{section}}
\renewcommand{\theequation}{\thesection.\arabic{equation}}

\parindent0em


\begin{titlepage}
\begin{center}
\hfill HUB-EP-97/17\\
\hfill {\tt hep-th/9703007}\\

\vskip .7in

{\bf  A CLASS OF $N=1$ DUAL STRING PAIRS AND ITS MODULAR SUPERPOTENTIAL}

\vskip .3in

{\bf Gottfried Curio and Dieter L\"ust}\footnote{email: \tt 
curio@qft2.physik.hu-berlin.de,			
		luest@qft1.physik.hu-berlin.de}
\\
\vskip 1.2cm

{\em Humboldt-Universit\"at zu Berlin,
Institut f\"ur Physik, 
D-10115 Berlin, Germany}

\vskip .1in

\end{center}

\vskip .2in

\begin{center} {\bf ABSTRACT } \end{center}
\begin{quotation}\noindent

We compare the $N=1$ F-theory compactification of Donagi, Grassi 
and Witten with modular superpotential - and some closely related models -
to dual heterotic models. 
We read of the $F$-theory spectrum from the cohomology
of the fourfold and discuss on the heterotic side 
the gauge bundle moduli sector 
(including the spectral surface) and the necessary fivebranes. Then we 
consider the $N=1$ superpotential and show how a heterotic superpotential 
matching the F-theory computation is built up by worldsheet instantons.
Finally we discuss how the original modular superpotential 
should be corrected by an additional modular correction factor, 
which on the $F$-theory side matches 
nicely with a `curve counting function' for the del Pezzo surface.
On the heterotic side we derive the same factor demanding correct
$T$-duality transformation properties of the superpotential.

\end{quotation}
\end{titlepage}
\vfill
\eject

\newpage

\section{Introduction}

During the last two years
accumulating 
and convincing evidence for the $N=2$ string-string duality
between heterotic string on $K3\times T^2$ and the type IIA string on a 
corresponding Calabi-Yau three-fold was obtained 
\cite{KV,FHSV,WKLL,KLT,KKLMV,AGNT,CCLMR}.
The $N=2$ string-string duality can be, at least heuristically, 
derived by
considering as a starting point 
the $N=4$ string-string duality \cite{DUFF}
between
the heterotic string on $T^6$ and the type IIA string on $K3\times T^2$
and then performing a kind of orbifoldization which breaks half of
the space-time supersymmetry.
The information about the (perturbative) 
heterotic spectra is encoded by a particular
choice of a gauge bundle over $K3$, which has to be matched by the
topological 
data of the type IIA $K3$-fibration.
Furthermore, non-perturbative states can emerge  when
considering various types of (compactified) branes on both sides.

The same type of techniques can be also applied when constructing
dual string pairs with $N=1$ supersymmetry in four dimensions.
Namely first, dual $N=1$ string pairs 
were obtained by orbifolding already known $N=2$ dual pairs 
\cite{VW}.
More
recently there has begun a
corresponding investigation of the $N=1$ duality between
the heterotic string on a Calabi-Yau three-fold - 
assumed to be elliptically fibered over a complex surface - 
together with a certain bundle embedded in the gauge bundle, 
and F-theory \cite{V} on a Calabi-Yau 
four-fold, which is assumed to be $K3$ fibered over the same surface, i.e.
one is adiabatically extending the corresponding 
eight-dimensional duality
\cite{W,DGW,GM,G,M}. In a certain sense we will combine both
techniques in this paper.

Now besides matching the spectra and enhanced gauge symmetries 
there were also refined checks of the $N=2$ duality, where 
a holomorphic quantity,  the prepotential,  
was compared on the heterotic side 
and on the type II side. 
There, one was restricted to weak coupling on the heterotic side,
where - because of T-duality - modular 
functions played an important role, 
whereas for the corresponding quantity on the 
type II side a world-sheet instanton sum played the dominant role. 

Investigating $N=1$ dual string pairs, possible checks, that go beyond
matching the spectrum, involve the comparison of $N=1$ 
effective interactions
which are determined by holomorphic quantities, namely
the superpotential or the gauge kinetic function.
In this paper we
will make a duality match between the superpotential, generated  by 
perturbative effects on the heterotic side, and on the $F$-theory
side
a certain sum over geometrical objects, which
produce instanton contributions (five-branes in the M-theory
set-up wrapped over certain six-cycels resp. the type IIB three-branes 
over corresponding four cycles \cite{W}). 
We will consider  models, where the F-theory four-folds 
are $K3$ fibrations over $dP$ of Euler number $\chi=12 \cdot 24$.
($dP$ stands for the del Pezzo surface $B_9$, the
projective plane blown up in the nine intersection points of two cubics.)
Moreover, these four-folds 
are elliptically
fibered over $dP\times P^1$ (the non Calabi-Yau three-fold base of type
IIB with varying dilaton).
The nice thing in this class of models \cite{DGW} is that, like as
for the $N=2$ prepotential, we will get for the
superpotential a modular function and 
furthermore the six-cycels reduce effectively again to rational curves in
the threefold $dP\times P^1$; namely the relevant
four-cycels are of the form `$\mbox{section}\times P^1$', 
where the first factor describes a section of the elliptic fibration of 
the del Pezzo over its own rational base $P^1_{dP}$.
The Calabi-Yau threefold of all
the heterotic dual models we are considering is given by an
elliptic fibration over $dP$, which has 
Hodge numbers $h^{(1,1)}=h^{(2,1)}=19$, denoted here as
$CY^{(19,19)}$ \cite{MV}.
The different $F$-theory compactification just correspond to different
choices of heterotic gauge bundles over $CY^{(19,19)}$.
We show that the rational curves on the heterotic
side reproduce the modular $F$-theory superpotential.

Our paper is organized as follows.
After discussing the four-fold $X^4$ which was used in \cite{DGW}
to obtain the modular superpotential, we will consider in chapter 
two closely
related $F$-theory compactifications which can be obtained from $N=2$
supersymmetric $F$-theory compactifications  by a $Z_2$ modding.
Specifically, the $N=2$ parent fourfolds will be either
given by $CY^{(3,243)}\times T^2$ (equivalent to IIA compactification
on $CY^{(3,243)}$), or by $K3\times K3$. In the first type of models
the 
non-vanishing Euler number 
of the $N=1$ fourfold $X^4$ and hence the  twelve three-branes  
emerge by
the ${\bf Z}_2$ modding; at the same time
the visible $dP$ emerges from a ${\bf Z}_2$
modding of $P^1\times T^2$; therefore we call this model 
of {\it `emergence'} type.
The second class of models,
in which one first goes to eight dimensions
and then to four dimensions on $K3$, we call {\it `reduction'} type since
the Euler number and so 
the number of 24 three-branes is reduced by half due
to the ${\bf Z}_2$ modding. Similarly the visible $dP$ is reduced
from the $K3$ by the modding procedure. 
On the heterotic side these two different types
of $F$-theory compactification will correspond to different choices
of gauge bundles with, however, same internal Calabi-Yau threefold
$CY^{(19,19)}$.
One can regard the different heterotic gauge bundles also  having
either a six-dimensional or eight-dimensional origin, respectively.
We will also discuss 
the spectral surface in the 
bundle moduli sector and the emergence/reduction of 
the corresponding twelve heterotic fivebranes. 
  
In chapter three we will discuss  
how to match the $F$-theory and heterotic superpotentials. 
  In this context the question arises of how to correct the 
  superpotential
  of \cite{DGW}
  by an $\eta$ power denominator, which is derived first via
  mirror symmetry and then independently via a heterotic orbifold 
  computation 
  using the modular weight arguments based on the fact that the 
  superpotential has  to balance the K\"ahler 
  potential with respect to $T$-duality transformations \cite{FLST}. 
  We also observe the occurence of a second $E_8$ theta-function.

For convenience of the reader some facts on the del Pezzo 
surface $dP={\scriptsize 
\left[\begin{array}{c|c}P^2&3\\P^1&1\end{array}\right]}$, 
the Calabi-Yau space $CY^{19,19}={\tiny 
\left[\begin{array}{c|cc}P^2&3&0\\P^1&1&1\\P^2&0&3\end{array}\right]}$ 
and the Calabi-Yau fourfold $X^4$ of \cite{DGW}, which are assumed to be 
known throughout the main body of the paper, are collected in the 
appendix.

\section{$F$-theory over $dP\times P^1$ and dual heterotic models on the 
$CY^{19,19}$}
\setcounter{equation}{0}

We start with a heuristic comparison of the incomplete data consisting of
the threefold base $B^3$ of the elliptically fibered Calabi-Yau fourfold 
on the F-theory side and the heterotic Calabi-Yau (without bundle).
Then we go on and describe (section 2.1) Calabi-Yau fourfolds 
elliptically 
fibered over the base $B^3=dP\times P^1$ and complete also (section 2.2)
the specification of the (0,2) heterotic Calabi-Yau model which involves 
additionally the choice of a stable, holomorphic vector bundle to be 
embedded in the gauge bundle.

The Calabi-Yau fourfold $X^4$ 
(we call it model A) for F-theory compactification used in \cite{DGW}
giving a modular superpotential is defined as
\beqa
X^4_A={\footnotesize 
\left[\begin{array}{c|cc}P^2_x&3&0\\P^1_y&1&1\\P^1_z&0&2\\P^2_w&0&3
\end{array}\right]}
\eeqa
representing a complete intersection in the product of the projective
spaces listed on the left given by two equations of the listed multidegrees.

Let us first discuss the fibration structure of the fourfold $X^4$.
Because of the two plane cubics occuring here $X^4$ can be seen in two ways
as being elliptically fibered over a threefold base. 
Especially  the elliptic F-theory fibration by the $T^2_w$ 
of the last row over $B=dP_{x,y}\times P^1_z$ is characteristic
for  model A.
Obviously $X^4_A$ is a $K3={\scriptsize 
\left[\begin{array}{c|c}P^1_z&2\\P^2_w&3\end{array}\right]}$
fibration over the mentioned del Pezzo
(more precisely, the $K3$   varies over the base $P^1_y$ 
of the $dP$, but not over its elliptic fibre $T^2_x$. Its Picard number is 
$\rho=2$, leaving 18 deformations).
If you fibre the fourfold over $P^1_y$,
the threefold fibre is given by $T^2_x\times K3$; this exhibits
the total space as the fibre product $X^4_A=dP\times _{P^1_y}{\cal B}_A$ 
of Euler number $\chi=12\cdot 24$ 
with the non-CY three-fold ${\cal B}_A={\tiny 
\left[\begin{array}{c|c}P^1_y&1\\P^1_z&2\\P^2_w&3\end{array}\right]}$,
which is fibered by the mentioned K3 over $P^1_y$. 

Let us now determine the heterotic Calabi-Yau 3-fold 
which is dual to $X^4_A$
or better to F-theory over $dP\times P^1$.
As the $X^4$ models lead to type IIB on $B=dP\times P^1_z$ one can use the 
duality in eight dimensions between type IIB on $P^1_z$ 
(resp. - taking into account the additional information provided by the 
7-brane locations/degenerate elliptic fibers - between F-theory on the 
$K3={\scriptsize 
\left[\begin{array}{c|c}P^1_z&2\\P^2_w&3\end{array}\right]}$) and the 
heterotic string on $T^2_{het}$ and then spread it out over $dP$ to four 
dimensions. The volume of $P^1_z$ will correspond to the heterotic dilaton.
This leads to the heterotic string on a Calabi-Yau threefold, 
which is elliptically fibered over del Pezzo,
for which purpose the $CY^{(19,19)}=dP\times _{P^1}dP$
presents itself naturally.\footnote{Note that besides the 
already visible $dP$ the second one can be argued heuristically to arise as 
follows: if one considers for the moment only this sector, i.e. (in 6D say)
the variation of $T^2_{het}$ over the
base $P^1_y$ of the visible del Pezzo, then this should match 
the variation
of the $K3_{12-8}$ over $P^1_y$ on the F-theory side, i.e.
the K3 fibered (non CY) threefold
${\cal B}_A={\tiny 
\left[\begin{array}{c|c}P^1_y&1\\P^1_z&2\\P^2_w&3\end{array}\right]}$;
but this space can be pulled back quadratically in the base $P^1_y$
to a CY just like the corresponding pull back would lead on the 
heterotic side
from del Pezzo to a 
$K3={\scriptsize 
\left[\begin{array}{c|c}P^1&2\\P^2&3\end{array}\right]}$
appropriate to correspond to a CY.}
Note, that besides being an elliptic fibration, $X^4_A$ itself is also a 
fibration of the $CY^{19,19}_{x,y,w}$ over $P^1_z$.\\
Let us remark that we will have some variability for 
the model building to follow: the only feature of $X^4_A$ 
which matters for the
modular superpotential is that it is a fibre product 
$dP\times _{P^1}{\cal B}$
built with a threefold ${\cal B}$
of $h^{1,0}({\cal B})=h^{2,0}({\cal B})=h^{3,0}({\cal B})=0$, 
which is K3 fibered over $P^1_y$ and even elliptically fibered over 
$F_0=P^1_y\times P^1_z$, so that the IIB base is $dP\times P^1_z$.\\

\subsection{F-theory side}

Before we enter the discussion about specific $F$-theory compactifications,
let us consider the question of determination of the spectrum in somewhat
more general terms (we will consider the brane sector later).
Besides the K\"ahler and complex structure parameters related to 
$h^{1,1}-2$ (not counting the unphysical zero-size F-theory elliptic fibre
as well as not counting the class corresponding to the heterotic dilaton) and
$h^{3,1}$ respectively, we have to take into account the contribution of 
$h^{2,1}$ giving in total $h^{1,1}-2+h^{2,1}+h^{3,1}$ parameters
which equals
$\frac{\chi}{6}-10+2h^{2,1}$ according to \cite{SVW}.
All these contributions divide themselves between 
chiral and vector multiplets (just as in the analogous 6D $N=2$ 
case \cite{MV})
according to whether or not they come from the threefold base $B^3$
of the F-theory elliptic fibration.
So we expect for the rank $v$ of the $N=1$ vector multiplets (unspecified
hodge numbers relate to $X^4$) (cf. also \cite{Mo})
\beqa
v=h^{1,1}-h^{1,1}(B^3)-1+h^{2,1}(B^3)\label{spectrumv} 
\eeqa
and for the number $c$ of $N=1$ neutral chiral 
(resp. anti-chiral) multiplets
\beqa
c&=&h^{1,1}(B^3)-1+h^{2,1}-h^{2,1}(B^3)+h^{3,1}\nonumber\\
 &=&h^{1,1}-2+h^{2,1}+h^{3,1}-v=\frac{\chi}{6}-10+2h^{2,1}-v.
 \label{spectrumc}
\eeqa
Now note that as our models are fibre products of del Pezzo 
and a K3 
fibered threefold ${\cal B}$ one has with $h^{(1,1)}=10+\rho=12$,
where $\rho$ denotes the Picard number of the K3 of the threefold,
that 
\beqa
v=\rho -2+h^{2,1}(B^3).\eeqa

Now 
in constructing 
specific $F$-theory fourfolds,
we make use of  the fact that the heterotic $CY^{19,19}$ is a ${\bf Z}_2$
orbifold of the space $K3\times T^2$, which represents the geometric
starting point for a heterotic $N=2$ compactification.
As explained in the appendix A.2 
the $dP$'s in the
$CY^{(19,19)}$ arise in two different ways,
namely either by {\it `emergence'} 
or by {\it `reduction'} from $K3\times T^2$
(the differences will show up again in two
different choices of heterotic gauge bundles).
Therefore on the $F$-theory side again two possible $Z_2$ moddings
present themselves naturally. First
$X^4$ might be obtained by modding out the 
corresponding ${\bf Z}_2$ involution on 
$T^2\times CY={\tiny 
\left[\begin{array}{c|cc}P^2_x&3&0\\P^1_y&0&2\\P^1_z&0&2\\P^2_w&0&3
\end{array}\right]}$ (model A),
i.e. the model is a $Z_2$ orbifold of the type IIA string on the 
$CY={\tiny \left[\begin{array}{c|c}P^1&2\\P^1&2\\P^2&3\end{array}\right]}$.
On the other hand,  
$X^4$ can also be obtained by modding out ${\bf Z}_2$ involutions on
$K3\times K3={\tiny  
\left[\begin{array}{c|cc}P^2_x&3&0\\P^1_y&2&0\\P^1_z&0&2\\P^2_w&0&3
\end{array}\right]}$, which we call model C (we leave out model B to avoid
confusion in notations).

Let us first consider the {\it `emergence'} type of models, where the Euler
number, $\chi=12\cdot 24$, 
the three-branes, $n_3=12$, and also the $dP$ emerge after the $Z_2$ modding.
Note  that the Calabi-Yau 
${\tiny \left[\begin{array}{c|c}P^1&2\\P^1&2\\P^2&3\end{array}\right]}$ 
is elliptic over $F_0$ and
of Hodge numbers $h^{(1,1)}=3$ as coming from the factors of the ambient
space and $h^{(2,1)}=3\times 3\times 10-(3+3+8)-1=75$
(for more on this crucial number 75 cf. appendix A.3); this does not 
satisfy
the six-dimensional anomaly condition for an elliptically fibered Calabi-Yau
threefold of $h^{(1,1)}=3$, which would be forced to have $h^{(2,1)}=243$.
But note that this CY, as well as the fourfold $X^4_A$,
does not have a section (as its K3 
already only has a 
trisection: the line in $P^2$).
So we will postpone the discussion of this model to the appendix. It may
still exist as a genuine (not modded from a $N=2$ situation) $N=1$ model.

So we will use instead of $CY^{(3,75)}$ the `consistent' $CY^{(3,243)}$
(of equation $y^2=x^3-f_{8,8}x-g_{12,12}$ with $h^{(2,1)}=9^2+13^2-(3+3+1)$),
i.e. we actually will consider the model $A^{\prime}$
\beqa 
X^4_{A^{\prime}}=(T^2\times CY^{(3,243)})/{\bf Z}_2=dP\times _{P^1_y}{\cal B}_{
A^{\prime}},
\eeqa
where ${\cal B}_{A^{\prime}}$ is the appropriately ${\bf Z}_2$
modded $CY^{(3,243)}$ , i.e. ${\cal B}_{A^{\prime}}:y^2
=x^3-f_{4,8}x-g_{6,12}$ with
$5\times 9+7\times 13-(3+3+1)=129$ deformations,\footnote{whereas 
$h^{2,1}({\cal B}_{A^{\prime}})=112$ as we will see 
(compare the corresponding difference of $\sharp def {\cal B}_A=45$,
$h^{2,1}({\cal B}_A)=28$ in the $A$ model (cf. A.3)); 
note that the number of complex 
deformations $\sharp def{\cal B}$ of the non-Calabi-Yau space ${\cal B}$
differs from $h^{2,1}({\cal B})$ by 17, resp. $\sharp def K3_{12-8}-1$ in
general, which equivalently makes possible to have the identity 
$h^{2,1}({\cal B})=h^{2,1}(X^4)$ as 
$\sharp def{\cal B}=\sharp def K3-1+h^{2,1}({\cal B})=
20-\rho-1+h^{2,1}({\cal B})\leftrightarrow h^{3,1}=8+3+\sharp def{\cal B}=
30-\rho+h^{2,1}({\cal B})\leftrightarrow h^{2,1}=h^{1,1}+h^{3,1}-40=
h^{2,1}({\cal B})$}
which then gives 
$h^{(3,1)}(X^4_{A^{\prime}})=8+3+129=140=5\times 28$ and using 
$40=\frac{\chi}{6}-8=h^{1,1}-h^{2,1}+h^{3,1}$ (cf. \cite{SVW}) and 
$h^{1,1}=12$ that
$h^{(2,1)}(X^4_{A^{\prime}})=112=4\times 28$.

Having computed all the relevant Hodge numbers we can easily determine
the spectrum for model $A^{\prime}$ from eqs.(\ref{spectrumv}) and
(\ref{spectrumc}).
Recall that 
$h^{2,1}(B^3=dP\times P^1)=0$ and $\rho=2$,
we  finally derive that $v=0$ and ($\chi=288$)
\beqa
c=38+2h^{2,1}=262.
\eeqa

Now let us come to the {\it `reduction'} type models, where the
Euler number, $\chi=12\cdot 24$,
and the three-branes, $n_3=12$, are 
obtained via reduction by the $Z_2$ modding.
Moreover here $K3$ is reduced to $dP$.
Using
$X^4_{C,C^{\prime}}=(K3\times K3)/Z_2$ with the $(10,8,0)$ 
involution (cf. \cite{B}) in the (say) first $K3_{8-4}$ 
giving the visible del Pezzo we consider the two options 
$(10,10,0)$ resp. $(10,8,0)$ concerning
the involution of Nikulin type $(r_2,a_2,\delta)$ in the 
second $K3_{12-8}$.
Note that these spaces are still fibre products of del Pezzo
(coming from $K3_{8-4}$) and a threefold ($K3_{12-8}$ fibered over the $P^1$ 
base of del Pezzo) over the $P^1$ base of del Pezzo.
One gets $h^{(1,1)}=10+r_2+\alpha$, where $r_2=\rho=10$ and $\alpha$
denotes the contribution from resolving the singularities caused by the
fixed locus in case $C^{\prime}$. 
Furthermore
$h^{(2,1)}_C=0$ resp. $h^{(2,1)}_{C^{\prime}}=8$ as this odd cohomology 
can come only from the fix locus: 
the two K3 lead to two base $P^1$ and two elliptic 
directions and in the case $C^{\prime}$ one has $2\times 2=4$ ordinary 
${\bf Z}_2$ singularities (in the `plane' built by the two $P^1$ directions)
`multiplied' by the two elliptic directions, which leads for each of the four
loci to $P^1_{\mbox{res}}\times E_{\mbox{vis}}\times E_{11,12}$ of 
respectively four $h^{(2,1)}$ classes (by wedging in the mentioned order
the classes in $(h^{1,1},h^{1,0},h^{0,0}),(h^{1,1},h^{0,0},h^{1,0}),
(h^{0,0},h^{1,1},h^{1,0}),(h^{0,0},h^{1,0},h^{1,1})$) of which only
the first two lead to new cohomology in $X^4$.
So one gets that 
for $((10,8,0),(10,10,0))$ of base 
     $dP\times P^1=dP\times _{P^1_y}F_0$, i.e. model $C$,
   $h^{(1,1)}=20$, $h^{(2,1)}=0$, $h^{(3,1)}=20$, so $v=8$,
  corresponding to a rank 8 gauge group, and $c=38-8=30$.
    
 For model $C^{\prime}$    
with $((10,8,0),(10,8,0))$ involution of base 
  $dP\times _{P^1_y}Bl_4(F_0)$ we derive
  that $h^{(1,1)}=24$, $h^{(2,1)}=8$, $h^{(3,1)}=24$ so $v=8+4$, 
  $c=38+2\cdot 8-(8+4)=42=30+12$.\footnote{We expect that
  the 4 vector multiplets and 12 chiral multiplets, which come in addition
  compared to model $C$, are non-perturbative on the
  heterotic side, since they
  arise from the blowing up of the type II base,
  like four additional heterotic fivebranes (wrapping now $T^2_{56}$
  instead of $T^2_{9,10}$) with their $12=4\cdot 3$ parameters.} Note 
  that the newly introduced classes in the $Bl_4$ process do not lead 
  to a further divisor contributing to the superpotential as 
  $\chi_{\mbox{ar}}(P^1_{\mbox{res}}\times E_{\mbox{vis}}\times E_{11,12})=
  0\neq 1$ because this divisor has $h^{3,0}=0$, $h^{2,0}=1$, $h^{1,0}=2$.

\vskip0.5cm

\subsection{Heterotic side}

The nice thing about the $CY^{(19,19)}$ is of 
course that it is a ${\bf Z}_2$
orbifold of $K3\times T^2$.
Now a $N=2$ heterotic string model on $K3\times T^2$ is 
specified by a choice
of gauge bundle in $E_8\times E_8$.  
If we consider a $(n_1,n_2;n_5)$ situation, where besides
an $SU(2)$ gauge bundle with instanton numbers $(n_1,n_2)$
in $E_8\times E_8$ also $n_5$ heterotic fivebranes are turned on,
then the anomaly cancellation condition in six dimension reads
$n_1+n_2+n_5=24$.
Specifically we are considering the complete Higgsed situation
which is equivalent to start from $E_8\times E_8$ instantons. 
Models $A^{\prime}$ and $C$ will represent the extreme choices of numbers
of heterotic five branes, namely $n_5=0$ or $n_5=24$ respectively,

Specifically, the $N=2$ parent of the heterotic dual 
of model $A^{\prime}$ is characterized by $n_1=n_2=12$, $n_5=0$.
After the ${\bf Z}_2$ modding, breaking $N=2$ to $N=1$,
the number of moduli is $h^{(1,1)}_{het}+h^{(2,1)}_{het}+x$,
where $x$ denotes the number of heterotic gauge bundle parameters,
i.e. here the number of surviving instanton moduli. 
Note that
the absence of fivebranes, $n_5=0$, in 
the (12,12;0) situation is consistent 
with the absence of 3-branes in 
the $N=2$ $F$-theory, since the Euler number of $CY^{(3,243)}\times T^2$ 
is zero. As in this case there are no preexistent
fivebranes let us see how after going to $N=1$
the (necessary to match the $12=\frac{\chi}{24}$ F-theory threebranes) 
heterotic fivebranes arise by {\it `emergence'}. Namely one has to fulfill 
$c_2(V_1)+c_2(V_2)+n_5 [f]=c_2(CY)$, where a number $n_5$ of fivebranes
wrapping the elliptic fibre f of the CY over its base B is allowed
(and required).
The evaluation $c_2(CY)\cdot J_1=
3\cdot 3+3\cdot 9=36$ then gives (via the relation of $J_1$ with $f$) 
the relation $n_5=12$ (cf. also \cite{FMW},\cite{BJPS} and A.2).   

Let us now come to the discussion about the heterotic spectrum, 
in particular the question about the heterotic gauge bundle.
Since the gauge group was completely broken by the (12,12) instantons
we could expect therefore that after the $Z_2$ modding 
(which acts freely on the original six-torus, see appendix A.2)
there are no $N=1$ vector multiplets, in agreement with the $F$-theory
prediction $v=0$.
Next consider the surviving scalar fields after the $Z_2$ twist.
Recall that the number of $N=2$ hypermultiplets was given by the
number of $K3$ deformations plus the quaternionic dimension 
of the  instanton moduli space,
\beqa
m_{inst}=\dim_{quat}({\cal M}^{inst}_{12}\times {\cal M}^{inst}_{12})=
2(c_2(E_8)\times 12-248)=2\times 112.
\eeqa 
Hence in the $N=2$ situation 
we count $H=20_{K3_{het}}+ m^{N=2}_{inst}=244$
hypermultiplets.
After $Z_2$ modding we get as number of chiral deformations first the
number of K\"ahler and complex structure parameters of $CY^{(19,19)}$,
i.e. $h^{1,1}_{het}+h^{2,1}_{het}=19+19$.
Second, of each of the $N=2$ instanton hypermultiplets there survives
one of their two chiral multiplets.
 So in total
\beqa
c&=&h^{1,1}_{het}+h^{2,1}_{het}+x \\
 &=&19+19+m_{inst} \\
 &=&38+224.
\eeqa
This number  matches\footnote{Note that in the setup 
of such a ${\bf Z}_2$ modding of type IIA on $CY^3$ (i.e.
model $A^{\prime}$) one has with a 
number of
$\sharp H=h^{2,1}(CY^3)+1$ hypermultiplets, 
$e_{CY^3}=2e_{{\cal B}}-2\cdot 24$
by the ramified covering and $e_{CY^3}=2(3-h^{2,1}(CY^3))$, 
$e_{{\cal B}}=
2+2(3-h^{2,1}({\cal B}))$ as $h^{3,0}({\cal B})=0$ 
the conclusion $2h^{2,1}({\cal B})=m^{N=2}_{inst}$
(so $h^{2,1}({\cal B})=112$) which 
gives with $h^{2,1}(X^4)=h^{2,1}({\cal B})$
the match $2h^{2,1}(X^4)=m^{N=2}_{inst}=x$.}
with the corresponding $F$-theory
prediction, i.e. $x=2h^{2,1}(X^4)$.

Let us indicate from a somewhat more general perspective that 
here - in the 
gauge bundle moduli sector - indeed the spectrum of the 
modded $N=2$ parent
model is simply the modded spectrum of the $N=2$ model.
For this note that in 
(leaving out the intermediate step over $X_{11}(j)$)
\beqa
\begin{array}{ccc} K3_{10-6}\times T^2_{6-4}&
\longrightarrow &CY^{19,19} \\
\;\;\;\;\;\;\;\;\;\;\;\;\;\downarrow -ell_{K3}& &
\;\;\;\;\;\;\;\;\;\;\;\downarrow -ell_{dP_{10-6}}\\
P^1_{K3}\times T^2_{6-4}&
\longrightarrow &dP_{8-4} \end{array}
\eeqa
a bundle V over $CY^{19,19}$  - to 
be considered as being modded from the $N=2$
situation on the left hand side - 
has by consistency to pullback to a bundle
'living' (varying) purely in the K3-sector, i.e. V must not vary along
$ell_{dP_{8-4}}$. So the spectral surface $C^2_{spec}$ 
(at first over $X_{11}$ in the intermediate step, say)
is double
covered (with branching only in codimension two) by
$C^1_{spec}\times T^2_{6-4}$ with $C^1_{spec}$ the
spectral curve of the $N=2$ parent model on the left.
So counting the deformations in the spirit of \cite{BJPS}
one finds that
again by $h^{2,0}(C^2_{spec})=h^{2,0}(C^1_{spec}\times T^2)=
h^{1,0}(C^1_{spec})$ the relevant number simply persists. 

Furthermore this sheds in our special case also light on a conjectured
relation \cite{FMW} between $h^{2,1}(X^4)$ and $h^{1,0}(C^2_{spec})$:
$h^{2,1}(X^4)=h^{2,1}({\cal B})=m^{N=2}_{inst}$ was the relevant
number of deformations on the $N=2$ level, i.e. $h^{1,0}(C^1_{spec})$,
and on the other hand
$h^{1,0}(C^2_{spec})=h^{1,0}(C^1_{spec}
\times T^2)=h^{1,0}(C^1_{spec})+1$.
Concerning the also
conjectured relation between the abelian varieties, the Albanese
$Alb(C^2_{spec})$ and the intermediate Jacobian
$Jac_{intmed}^{(2,1)}(X^4)$, note that in our setup the first is
now  related to $Jac(C^1_{spec})$, whose relation with
$Jac_{intmed}({\cal B})$ (related to 
$Jac_{intmed}^{(2,1)}(X^4)$, extending their
dimensional identity; $Jac_{intmed}({\cal B})$ occures here as
capturing the relevant part of $Jac_{intmed}(CY^3_F)$)
should then be  part of a corresponding $N=2$ relation.

Let us, more briefly, also discuss the heterotic duals of the
$F$-theory models $C$ and $C^{\prime}$.
For model C we need on the level of the $N=2$ parent model 
$24=\frac{\chi_{K3\times K3}}{24}$ threebranes on the 
$F$-theory side; so we need $n_5=24$ on the heterotic side, and we
do not turn on any gauge bundle
in the dual heterotic model, i.e.  $(n_1,n_2,n_5)=(0,0;24)$. 
(The 
gauge group $E_8\times E_8$ would remain unbroken in six dimensions.)
After the ${\bf Z}_2$ modding the 12 heterotic fivebranes
arise by {\it {`reduction'}} from the 24 fivebranes in the $N=2$ situation.
Remember that for models C, C' we obtained $h^{(2,1)}(X^4_{C})=0$, 
$h^{(2,1)}(X^4_{C'})=8$.
Since on the heterotic side there are no instantons turned on, one
expects no gauge bundle parameters, but a surviving gauge group 
from the unbroken $N=2$ gauge group $U(1)^{16}$, which gives (in the C model,
say) a rank 8 group; furthermore the greater rigidity on the F-theory side
($\sharp def K3_{12-8}=20-\rho=10$ only instead of 18) translates itself
to a corresponding rigidity of the $CY^{19,19}$ freezing also 8 moduli, i.e.
leaving only $c=38-8=30$ moduli there.

\section{The Dual Superpotentials}

\subsection{$F$-theory superpotential}

Recall that
the authors of \cite{DGW} find for the $F$-theory on $X^4$ a superpotential
which is represented in the type IIB language by wrapping three branes
over the four cycles of the form $C\times P^1_z$, where C is a rational
curve in the del Pezzo of selfintersection $C^2=-1$ (a condition being 
equivalent for a rational $C$ on $dP$ to $C^2<0$ and furthermore to being
a section of the elliptic fibration), i.e.
\beqa
W=\sum_{C^2=-1,C\;\mbox{rational}}e^{2\pi i<c(C),z>}
\eeqa
up to a prefactor, common to all divisors, with dependence on the complex
structure moduli (there are further possibilities as well \cite{W5},\cite{G}).
Here $c(C)$ denotes the homology class of $C$, say expressed as\footnote{Here 
by abuse of 
language we do not distinguish between the curves and their homology classes}
$c_0F+\sum_{i=1}^9 c_iE_i$ where $F$ is the elliptic fibre of del Pezzo
and the $E_i$ are the nine blown up intersection points of two cubics in
the projective plane; $z=(z_i)_{i=0,...,9}$ is
a corresponding ten parameter vector in the dual cohomology, i.e.
$<c(C),z>=\sum_{i=0}^9 c_iz_i$. If one changes to base systems adapted to
the $E_8$ intersection lattice\footnote{Essentially $A_i=E_i-E_{i+1}$, where
a $C$ then has coordinates $m_i$, say, instead of the $n_i$, and 
$w_i=z_i-z_{i+1}$, $i=1,...,7$; cf. \cite{DGW} for details; 
$z_9$ is the K\"ahler modulus of the
the base $P^1_y$; $\tau$ is the K\"ahler modulus of the fibre 
$T^1_x$, and the $w_i$  correspond to the $E_8$ part;
note that 
$m^2:=\sum_{i=1}^8 m_i^2-(m_1m_2+\cdots +m_6m_7+m_3m_8)$ is $-\frac{1}{2}$
times the $\vec{m}^2$, say, built with the intersectionform of the (negative
definite) $E_8$ intersection lattice} one has with $z_0:=\tau$ for such a C 
that $<c(C),z>=c_0z_0+\sum_{i=1}^9 c_iz_i=
(m^2+\frac{m_8}{3})\tau+\sum_{i=1}^8 m_iw_i+(z_9-\frac{m_8}{3}\tau)=
m^2\tau+(m,w)+z_9$, so ($q_9=e^{2\pi i z_9}$):
\beqa
W=q_9 \Theta _{E_8}(\tau,w_i)
 =q_9\sum_{m\in {\bf Z}^8_{E_8}}q_{\tau}^{m^2}\prod_{i=1}^8q_{w_i}^{m_i}\sim
  q_9\sum_{a=1}^4\prod_{i=1}^8\th_a(\tau,w_i),\label{supwi}
\eeqa
being of modular weight 4 with respect to 
$PSL(2,Z)_\tau$ (the $w_i$ transform as 
$w_i\rightarrow{w_i\over c\tau+d}$).
This superpotential is common to our models.
Minimizing this superpotential leads to a supersymmetry preserving locus 
(essentially unique, i.e. up to the action of the Weyl group of $E_8$)
consisting in locking pairs of the $w_i$ on the four half-periods of the 
elliptic curve $E_{\tau}$. Expanding in $\phi_i=w_i-w_i^0$ 
around the minima $w_i^0$ 
gives $W|_{SUSY}\sim q_9\th_1^2(\tau,\phi)\eta(\tau)^6$
behaving in leading order as $q_9\phi^2\eta(\tau)^{12}$
(for notational simplicity we have identified all $\phi_i$).

Now we will give some arguments that
the superpotential eq.(\ref{supwi}) has 
to be corrected by a modular function. In fact, the 
authors of \cite{DGW} expect that 
this expression for the superpotential 
has to be corrected by an $\eta^8$ denominator - leading to a completely 
modular invariant superpotential - when taking into account a correct 
counting of the sum of rational (-1)-curves including also reducible 
objects.
We would like to argue that a different correction factor is
required to get the correct modular weight for $W$,
namely a factor $\eta(\tau)^{-12}$. Note that then
the corrected superpotential $W'=W/\eta^{12}$ of modular weight $-2$ is
around the minima $w_i^0$ simply given by a $\tau$-independent
mass term ($\mu$-term) for the fields $\phi$ 
\beqa
W^{\prime}|_{SUSY}\sim q_9\phi^2.\label{wprimesusy}
\eeqa
To argue for this correction by $\eta^{-12}$
we can compare with a precise rational curve counting on the 
del Pezzo provided by mirror symmetry \cite{KMV}. 
In the $CY^{3,243}$ over $F_1$ (where $dP$ occurs over the exceptional
section of the $F_1$ base)
one finds among the instanton numbers
\beqa
\sum n_{0,1,k}q^k=\frac{E_4}{q^{-1/2}\eta^{12}},
\eeqa
where the zero index indicates that we are considering the del Pezzo sector
of the threefold and the 1 indicates that inside the del Pezzo itself one has 
$C\cdot f=1$ as intersection with the elliptic fibre of del Pezzo;
i.e. considering the $n_{0,1,k}$ sector implements exactly the counting we
want to do. Now in the del Pezzo lattice $H\oplus E_8=b{\bf C}\oplus
f{\bf C}\oplus E_8$ (after suitable base change; $b^2=-1,f^2=0,b\cdot f=1$)
one has for such a $C=\alpha b+\beta f+\vec{l}\cdot\vec{e}$ with $\alpha=1$,
$\beta=k$ that $-2=C^2+C\cdot K=C^2-C\cdot f=-1+2\beta +\vec{l}^2-1$, i.e. 
$\beta=-\frac{1}{2}\vec{l}^2$ showing the $E_4$ of the naive count.
So if we compare with the superpotential  
having all $w_i$ locked to zero, which is given by 
$W(\tau,w_i=0)=q_9E_4(\tau)$, we find the asserted correction factor.

Note that the factor $\eta^{-12}$ furthermore comes up not only also in a
$\eta^{-\chi}$ computation for del Pezzo, but even in\footnote{Namely 
$u=1-2\frac{\theta_2^4}{\theta_3^4}=\frac{\theta_4^4-\theta_2^4}{\theta_3^4}$ 
gives (using that 
$\theta_{3\mbox{\, resp.}2}^4=8\frac{1}{2\pi i}\partial _{\tau}\log
\frac{\theta_{2\mbox{\, resp.}3}}{\theta_4}$ (cf. \cite{DK}) so 
that $4\partial _{\tau}\log \frac{\theta_2}{\theta_3}=\pi i \theta_4^4$) that
$\partial _{\tau}u=(u-1)\partial _{\tau}\log (1-u)=
(u-1)4\partial _{\tau}\log \frac{\theta_2}{\theta_3}=(u-1)\pi i \theta_4^4=
-2\pi i\frac{\theta_2^4\theta_4^4}{\theta_3^4}$, so 
$\frac{\partial \tau}{\partial u}=
-\frac{1}{2\pi i}\frac{\theta_3^4}{\theta_2^4\theta_4^4}$; on the other hand
$u^2-1=\frac{(\theta_4^4-\theta_2^4)^2-\theta_3^8}{\theta_3^8}=
-4\frac{\theta_2^4\theta_4^4}{\theta_3^8}$ and so 
$(u^2-1)\frac{\partial \tau}{\partial u}=
\frac{4}{2\pi i}\frac{1}{\theta_3^4}$.}
\beqa
e^{b\chi+c\sigma}=[(u^2-1)\frac{d\tau}{du}]^{\chi/4}(u^2-1)^{\sigma/8}
=(\frac{2}{\pi i})^3\frac{1}{\theta_3^{12}}
\frac{-\theta_3^8}{4\theta_2^4\theta_4^4}=
-(\frac{1}{2\pi i})^3\frac{1}{\eta^{12}}
\eeqa
occuring in connection with the question of integration over the u-plane
\cite{W-u} .

\subsection{Heterotic Superpotential}

Now according to \cite{W} the superpotential generating divisors
on the F-theory side correspond in our case to world-sheet instantons
on the heterotic side.
We want all the world-sheet-instantons/rational
curves to contribute to get a match with $F$-theory. 
So either we should not have a nontrivial
bundle embedded at all, i.e. we should reach our situation  in the 
bundle sector in models $C$, $C^{\prime}$, i.e. 
from an $(0,0;24)$ startpoint (anomaly cancellation 
purely by fivebranes) in $N=2$; 
so in this case the rational curves are not obstructed
at all to contribute to the superpotential as the spin bundle
${\cal O}(-1)$ of a rational curve
 will not be tensored with an embedded bundle and so
no fermion modes are created.
 Alternatively 
we could start from
a nonstandard embedding (i.e. not purely in one $E_8$) like (12,12;0) in
$N=2$, i.e $A^{\prime}$ model setup, where also 
all world-sheet-instantons may contribute to
the superpotential.

The rational instanton  numbers of the $CY^{19,19}$ are 
essentially determined
by the $dP$ geometry (for more details see appendix A.2).
(Since the $dP$ base is common to the $F$-theory fourfolds and 
to the heterotic
$CY^{19,19}$ one can more or less immediately deduce the equality of
the superpotentials.)
So let us read of the (naively counted) 
rational instanton numbers  of the $CY^{19,19}$:  
\beqa
n_{k_{\tau^{\prime}},(k_{w_i^{\prime}}),k_9,k_{\tau},(k_{w_i})}=
\delta_{k_{\tau^{\prime}},(k_{w_i^{\prime}})^2}\delta_{{k_9},1}
\delta_{k_{\tau},(k_{w_i})^2}.
\eeqa
These instanton numbers lead to the following heterotic  
superpotential
\beqa
W=\Theta_{E_8}(\tau^{\prime},w_i^{\prime})q_9
\Theta_{E_8}(\tau,w_i).
\eeqa
Note that the heterotic computation leads to a second 
$E_8$ theta-function,
which should appear in the prefactor on the $F$-theory side.

Let us now discuss the modular properties of 
the heterotic superpotential.
If we follow the 
construction of the heterotic string on the Calabi-Yau $CY^{19,19}$ as an 
${\bf Z}_2\times {\bf Z}_2'$ orbifold (cf. A2), we can see 
that the perturbative heterotic superpotential, which describes
a mass term for Wilson line moduli fields, 
is constrained by the unbroken target space duality
symmetries of the orbifold in such a way that
the superpotential, which includes
the factor $\eta(\tau)^{-12}$, has the correct modular weight.
For this consider the $\tau$-dependent superpotential in the orbifold limit.
In $N=1$ supergravity the K\"ahler potential $K$ and the superpotential
$W$ are connected, and the matter part of the $N=1$ supergravity Langrangian
\cite{CREMMER} is described by a single  function
$G(\phi,\bar \phi)=K(\phi,\bar \phi)+\log|W(\phi )|^2$,
where the $\phi$'s are chiral superfields.
The target space duality  transformations act as discrete
reparametrization on the scalars $\phi$ and induce simultaneously a
 K\"ahler
transformation on $K$. Invariance
of the effective action constrains $W$ to transform as a 
modular form of particular weight \cite{FLST};
specifically under $PSL(2,Z)_{T_1}\times PSL(2,Z)_{T_3}$ the 
superpotential 
must have modular weights -1, i.e. it has to transform under
$T_{1,3}\rightarrow{a_{1,3}T_{1,3}+b_{1,3}\over c_{1,3}T_{1,3}+d_{1,3}}$ 
as
$W\rightarrow {W\over (c_{1}T_{1}+d_{1})(c_3T_3+d_3)}$; 
in particular the $\mu$-term 
\beqa
W=\phi_{1,i}\phi_{3,i}\label{muterm}
\eeqa
has the required modular weight 
and precisely matches with $W^{\prime}|_{SUSY}$
in eq.(\ref{wprimesusy})
(a more general
form would be given by 
$W={\phi_{1,i}^{l_1}\phi_{3,i}^{l_3}\over \eta(T_1)^{-2l_1+2}
\eta(T_3)^{-2l_3+2}}$).
Now, 
as explained in appendix A2, the Calabi-Yau K\"ahler modulus $\tau$
corresponds in the orbifold limit to the diagonal deformation
$\tau=T_1=T_3$. Then,
concerning the transformation properties 
of the superpotential under the diagonal  modular
transformations $PSL(2,Z)_\tau$,
invariance of the $G$-function requires that $W$ has modular weight -2, i.e.
that under
$\tau\rightarrow{a\tau+b\over c\tau+d}$ one has 
$W\rightarrow{W\over (c\tau+d)^2}$.
Clearly the mentioned $\mu$-term has the correct modular
weight (a more general function of $\tau$ and $\phi_i$ is given by
$W={\phi_i^n\over\eta(\tau)^{-2n+4}}$). As already discussed, the 
superpotential (\ref{muterm}) has the
supersymmetry preserving minimum ($W=0$, $dW=0$) 
$\phi_{1,i}=\phi_{3,i}=0$.
Therefore  the vacuum expectation values of the Wilson line
fields $\phi_{1,i}$, $\phi_{3,i}$ are set to zero after 
the minimization, i.e. the vacuum expectation values are not free,
continuous parameters in the presence of this superpotential. 
Going away from the minimum of the superpotential by turning on
the Wilson line fields $\phi_{1,i}$, $\phi_{3,i}$ means in the context
of conformal field theory, that one is in fact going away from the
conformal point, i.e. going off-shell.

Finally let us also remark on the
factor $q_9$ in eq.(\ref{supwi}).
 In the orbifold limit the possible 
$z_9$-depedence of the superpotential is 
again restricted by $T$-duality.
However the duality group with respect to the modulus $z_9$ is
no longer the full modular group $PSL(2,Z)$ but only a subgroup of it,
since the $R\rightarrow 1/R$ duality is broken by the freely acting
$Z_2$ in this sector 
(the space $P^1$ has no $R\rightarrow 1/R$ duality due
to the absence of winding modes in this sector):
so the superpotential is not required to transform as a modular
function, but it should be just a periodic function in $\Re z_9$, like
$q_9$. These kind of functions generically arise as the zero mode
prefactor in the large $z_9$ limit (i.e. supressing all winding modes
in this decompactification limit) of modular functions, 
like the $\eta$-function or the $\theta$-functions.
Just consider the following naive example.
We can regard  \cite{FKLZ}
the superpotential 
as the sum over the massive (BPS) spectrum of the orbifold 
compactification: $W\sim\prod M^{-1}$. For example, summing over
all momentum and winding states of a two-torus compactifiaction with
masses $M=m+nT$ in a $SL(2,Z)_T$ invariant way yields the $T$-dependent 
superpotential
$W\sim \prod_{m,n}(m+nT)\sim \eta(T)^{-2}$, which has the required 
modular
weight -1.
Similarly summing over all momentum and winding modes of
the shifted lattice, corresponding to the free plane
$z_2$, one obtains\footnote{We
thank C. Kounnas for discussion on this point.}
a factor $W\sim\theta_2(T_2)^{-2}$, leaving one in
the limit of large $\Im T_2$, i.e. supressing all winding modes
in this sum, with the zero mode piece: 
$W\sim e^{-2\pi i T_2/ 4}=e^{-2\pi i z_9}=q_9^{-1}$.

In summary we have supported some strong evidence that the perturbative
heterotic superpotential matches with its $F$-theory counterpart.
It would be very interesting to analyze models with (modular) non-perturbative,
$S$-dependent 
heterotic superpotentials \cite{FILQ}  and their $F$-theory duals.
In this context non-perturbative symmetries in underlying $N=2$ models,
like the $S$-$T$ exchange symmetry in the $CY^{3,243}$, may play
an important role.

\vskip0.5cm

{\bf Acknowlegement}
We would like to thank B. Andreas, P. Aspinwall, M. Bershadsky, S. Kachru,
J. Louis,
N. Nekrasov, Y. Oz, R. Plesser, S. Theisen
 and especially G. Lopes Cardoso, C. Kounnas,
D. Morrison and E. Witten for useful discussions.
The work is supported by the Deutsche Forschungsgemeinschaft (DFG)
and by the European Commission TMR programme ERBFMRX-CT96-0045.

\vskip0.5cm

\appendix


\section{Appendix}

\subsection{The del Pezzo surface}

The representation 
${\scriptsize \left[\begin{array}{c|c}P^2_x&3\\P^1_y&1\end{array}\right]}$ 
of the del Pezzo makes 
visible on the one hand its elliptic fibration over $P^1_y$
via the projection onto the second factor; on the other hand the defining
equation $C(x_0,x_1,x_2)y_0+C^\prime (x_0,x_1,x_2)y_1=0$ shows that the 
projection onto the first factor exhibits $dP$ as being a $P^2_x$ blown up in
9 points (of $C\cap C^\prime$), so having as nontrivial hodge number (besides
$b_0,b_4$) only $h^{1,1}=1+9$. 
Furthermore the $dP$ has 8 complex structure moduli: they can be seen
as the parameter input in the construction of 
blowing up the plane in the 9 intersection points of two cubics
(the ninth of which is then always already determined as they sum up to
zero in the addition law on the elliptic curve;
so one ends up with $8\times 2-8$ parameters) or - counting via number of
inequivalent monomials - as $10\cdot 2-(8+3)-1$.\\
The $dP$ can be obtained from $K3$ by a ${\bf Z}_2$ modding.
This corresponds to having on K3 a Nikulin involution of type
(10,8,0) with two fixed elliptic fibers 
in the K3 leading to
\beqa
\begin{array}{ccc}K3&\rightarrow &dP\\\downarrow & &\downarrow\\P^1_y&
\rightarrow &P^1_{\tilde{y}}\end{array}
\eeqa
induced from the quadratic base map $y\rightarrow \tilde{y}:=y^2$ with the 
two branch points $0$ and $\infty$ (being the identity along the fibers).
One can follow this relation also in the orbifold representation 
of K3 as $T^4/Z_2$, 
where the involution operates on the $T^2$'s as sign-flip;
this shows also the fibration by the first, say, $T^2$ over the $P^1=T^2/Z_2$
coming from the second $T^2$ in a double covering having 4  branch points 
leading to four $\tilde{D}_4=I_0^*$ fibers.
Now (del Pezzo being K3 divided by an involution having two fixed fibers)
do a second ${\bf Z}_2$ modding given by an involution of the base
coordinate $y$ together with an half lattice shift $1/2$:
$(x,y)\rightarrow (x,-y+1/2)$.
This destroys essentially half
of the cohomology of $K3$ leading to $H\oplus E_8$ as
intersection form of $dP$.\\
In the Weierstrass representation $y^2=x^3-f_8(u)x-g_{12}(u)$ of $K3$
the mentioned quadratic redefinition translates to the representation 
$y^2=x^3-f_4(u)x-g_6(u)$ of $dP$ (showing again the $8=5+7-3-1$ deformations). 
Repeated use will be made in the paper of the fact that 
del Pezzo can be 
reached, in the sense of turning on complex deformations, from the ${\bf Z}_2$
modded (via the mentioned quadratic base map, now with sign-flip in the fibers)
{\em constant} elliptic fibration over $P^1$
\beqa
\begin{array}{ccc}P^1\times T^2&\rightarrow &X_{11}(j)\\ 
\downarrow & &\downarrow\\P^1&\rightarrow &P^1\end{array} \;\; .
\eeqa
Here $X_{11}(j)$ is the (almost) constant fibration of elliptic curve of 
invariant j with two $\tilde{D}_4=I_0^{\ast}$ singular fibers over the two
branch points of the quadratic base map (cf. \cite{MP}).
One finds this degenerate del Pezzo on the boundary of the complex structure
moduli space at $f_4=ru^2,g_6=su^3$: $j=\frac{4r^3}{4r^3+27s^2}$, the
two singular fibers are at $u=0,\infty \in P^1$ and one sees the four 
sections (in the covering above given by the constant fibration 
exactly the sections
given by the halfdivision points of the $T^2$ survive the modding): they are
$(x,y)=(x_iu,0)$, where $x_i$ solves $x^3+rx+s=0$, besides the the zero section
given by the point at infinity.

\subsection{The Calabi-Yau $CY^{19,19}$}

The $CY^{(19,19)}={\tiny 
\left[\begin{array}{c|cc}P^2&3&0\\P^1&1&1\\P^2&0&3\end{array}\right]}=
dP\times _{P^1_y}dP$,
which is elliptically fibered over del Pezzo, can be obtained from
$T^2 \times K3={\tiny 
\left[\begin{array}{c|cc}P^2&3&0\\P^1&0&2\\P^2&0&3\end{array}\right]}$
by the Voisin-Borcea involution, which consists in the 'del Pezzo' involution
(type (10,8,0) in Nikulins classification) with two fixed elliptic fibers 
in the K3 combined with the 
usual "-"-involution with four fixed points in the $T^2$; this leads 
in the base to the relation mentioned in A1
and in the $P^1_y\times T^2$ 'plane' to the $X_{11}$ mentioned in A.1 and so to
the second del Pezzo . So here the symmetric `degree one' entries in the 
$P^1$ variables have a seemingly different origin: one by {\it `reduction'} 
(from 
two) and one by {\it `emergence'} (from zero).
There is of course only an appearent asymmetry in
the situation: the fibration in the $P^1_y\times T^2$ `plane' with the two 
singular $\tilde{D}_4=I_0^{\ast}$ fibers is ${\bf Z}_2$ covered by the 
orbifold limit of K3 
(instead of the smooth
$K3={\scriptsize \left[\begin{array}{c|c}P^1&2\\P^2&3\end{array}\right]}$)
with four $\tilde{D}_4=I_0^{\ast}$ fibers, i.e. a fully 
symmetric start could be done from 
$T^6/({\bf Z}_2\times {\bf Z}_2^{\prime})$. 
So, to elaborate on the construction of the heterotic string
on the $CY^{19,19}$ as an ${\bf Z}_2\times {\bf Z}_2^{\prime}$ orbifold,
consider first the heterotic string on the six-torus $T^2_1\otimes T^2_2
\otimes T^2_3$ with K\"ahler moduli $T_j$ 
($j=1,2,3$), and complex structure moduli $U_j$. 
The ${\bf Z}_2\times {\bf Z}_2^{\prime}$ acts on the three complex coordinates
$(z_1,z_2,z_3)$ as 
\beqa
\alpha:\qquad (z_1,z_2,z_3)\rightarrow (-z_1,-z_2,z_3),\nonumber \\
\beta: \qquad (z_1,z_2,z_3)\rightarrow (-z_1,z_2+{1\over 2},-z_3),\nonumber\\
\alpha\beta: \qquad (z_1,z_2,z_3)\rightarrow (z_1,-z_2+{1\over 2},-z_3) .
\eeqa
These three ${\bf Z}_2$ moddings define three $N=2$ subsectors in the total
Hilbert space. The first $\alpha$-modding acts non-freely; this modding
corresponds to the orbifold limit of the heterotic string on $K3_{1,2}\times
T^2_3$.
In the same way, the non-free $\alpha\beta$-modding corresponds to 
a $K3_{23}\times T^2_1$ compactification. So the situation is symmetric 
with respect to the
$\alpha$- and $\alpha\beta$-modding thus reflecting the fibre product 
structure of the $CY^{19,19}$. On the other hand,  the generator
$\beta$ acts freely on the original six-torus, so we call $z_2$ 
the free plane. Let us relate now
the  moduli of the orbifold compactification on 
$(T^2_1\otimes T^2_2\otimes T^2_3)/({\bf Z}_2\times {\bf Z}_2^{\prime})$ 
and the Calabi-Yau moduli. 
First compare the moduli $T_j$ with the three Calabi-Yau moduli $z_9$,
$\tau$ and $\tau'$. The free plane $z_2$ corresponds to the $P^1$ base 
of $dP$, so $4z_9=T_2$, where the factor
4 arises because the volume is reduced two times by half going from 
$T^2$ to $P^1_{K3}$ to $P^1_{dP}$.
On the other hand, since the $K3\times T^2$ compactification is
obtained by the $\alpha$-modding as well as by the $\alpha\beta$-modding, the 
modulus $T_1$ does not correspond, say, to the modulus $\tau$ of the elliptic 
fibre of $dP$, but the moduli $\tau$ and $\tau'$ correspond to certain 
linear combinations of orbifold states: $\tau$, say, to the deformation 
along the diagonal $\tau=T_1=T_3$, $\tau'$ to the orthogonal deformation 
(the symmetry between the $\alpha$- and $\alpha\beta$-modding  
enforces us to take these linear combinations). 
The same type of identification holds 
for the other moduli like the $w_i$ of the $CY^{19,19}$ in terms of
again identified orbifold Wilson line fields, i.e.  
$w_i-w_i^0=\phi_i=\phi_{1,i}=\phi_{3,i}$, where the  
$\phi_{1,i}$, $\phi_{3,i}$ ($i=1,\dots, n_1=n_3=8$) belong to the first
resp. third torus . The classical  K\"ahler potential of the fields 
$T_i$, $U_i$, $\phi_{1,i}$ and $\phi_{3,i}$ has the form
$K=-\sum_{j=1}^3\log\lbrack (T_j-\bar T_j)(U_j-\bar U_j)-
\sum_{i=1,j\neq 2}^{n_j}(\phi_{j,i}+\bar \phi_{j,i})^2\rbrack$ .
The unbroken T-duality group contains $PSL(2,Z)_{T_1}\times PSL(2,Z)_{T_3}$, 
which acts as 
$T_{1,3}\rightarrow{a_{1,3}T_{1,3}+b_{1,3}\over c_{1,3}T_{1,3}+d_{1,3}},
\quad \phi_{1,i}\rightarrow {\phi_{1,i}\over c_1T_1+d_1},\quad
\phi_{3,i}\rightarrow {\phi_{3,i}\over c_3T_3+d_3}$.
Hence under simultaneous modular transformations 
($a=a_1=a_3$, etc.) along the diagonal, $\tau=T_1=T_3$, 
the group $PSL(2,Z)_\tau$ has the action
$\tau \rightarrow {a\tau+b\over c\tau+d},\quad \phi_i
\rightarrow {\phi_i\over c\tau+d}$.

Let us now consider the rational curves in the
$CY^{(19,19)}$. Let ${\cal O}(2)\oplus{\cal O}(a_1)\oplus{\cal O}(a_2)$ 
be the splitting type of (the tangential bundle of CY over) the rational
curve C. Then ${\cal O}(2)\oplus{\cal O}(a_i)$ is the 
corresponding splitting type of the projected rational curve $C_i$ 
of normal bundle ${\cal O}(C^2_i)$ in the del
Pezzos $dP_i$ ($i=1,2$). From 
$-e_{C_i}=C_i^2+C_i\cdot K_{dP_i}=a_i-C_i\cdot f_i$
you see that $a_i\ge -1$ (cf. also \cite{W})
which together with $a_1+a_2=-2$ shows $a_i=-1$ (in other words: {\em all} 
rational curves lying in 
$CY^{19,19}$ project in the $dP_i$ factors to the {\em special} rational 
curves of selfintersection -1).
That is we get for the (naively read of) instanton numbers of the 
$CY^{(19,19)}$ (cf. sect. 3.1 and \cite{DGW})
\beqa
n_{k_{\tau^{\prime}},(k_{w_i^{\prime}}),k_9,k_{\tau},(k_{w_i})}=
\delta_{k_{\tau^{\prime}},(k_{w_i^{\prime}})^2}\delta_{{k_9},1}
\delta_{k_{\tau},(k_{w_i})^2}.
\eeqa
To rephrase the process of constructing rational curves: in the beginning
one has the base $P^1_y$; then one embeddes it in the del Pezzo base as
section; then to get really a curve in the threefold represented 
by the fibered product of the two del Pezzos one has to do 
the same process also for the other fibration direction giving the symmetrical
result indicated; as the processes are - besides the common
base - independent, one gets the second factor.

For use in the next section let us point out the existence of
a conifold transition to the CY 
${\scriptsize \left[\begin{array}{c|c}P^2&3\\P^2&3\end{array}\right]}$ 
of $h^{(1,1)}=2$
and $h^{(2,1)}=10\cdot 10-(8+8)-1=83$ and so of Euler number -162. If you
start from $CY^{(19,19)}={\tiny 
\left[\begin{array}{c|cc}P^2_a&3&0\\P^1_y&1&1\\P^2_b&0&3\end{array}\right]}$
with the equations $y_0Q_a+y_1R_a=0$ and $y_0S_b+y_1T_b=0$ one sees that the 
existence condition for y gives the special bicubic $Q_aT_b-R_aS_b=0$ and the
singular set (for it) $Q=R=S=T=0$ of 81 nodes; i.e. contract in the 
$CY^{(19,19)}$ say 81 $P^1$'s (coming from combining respectively 9 sections
in each del Pezzo of the fibre product) and then deform to a generic bicubic
(i.e. detune $83-19=64=8\cdot 8=(9-1)\cdot (9-1)$ parameters).

Note that if $J_i$ $(i=1,2,3)$ denote the
induced classes from the factors in $CY^{(19,19)}={\tiny 
\left[\begin{array}{c|cc}P^2&3&0\\P^1&1&1\\P^2&0&3\end{array}\right]}$, 
$m_a$
the dimensions of the factor spaces and $d_i^a$ $(i=1,2)$ the 
respective degrees of the two defining equations one has 
$c_2^{ab}=\frac{1}{2}(-\delta^{ab}(m_a+1)+\sum_{i=1}^2 d_i^ad_i^b)$, so
$c_2=3(J_1^2+J_3^2+J_2(J_1+J_3))$. Furthermore one has (as  
$dP={\scriptsize \left[\begin{array}{c|c}P^2&3\\P^1&1\end{array}\right]}$ has 
intersection form 
${\scriptsize \left(\begin{array}{cc}1&3\\3&0\end{array}\right)}$
for the divisors (line resp. point) induced from the factors) for the 
intersection numbers of the CY that $K^0=3J_1^2J_3+3J_1J_3^2+9J_1J_2J_3$.

\subsection{The Calabi-Yau four-fold $X^4_A$}

Remember that 4-fold $X^4$ (model A) of section (2.1) could be represented
as the fibre product $X^4=dP\times _{P^1_y}{\cal B}$, where
${\cal B}=
{\tiny \left[\begin{array}{c|c}P^1_y&1\\P^1&2\\P^2&3\end{array}\right]}$.
This gives $h^{(1,1)}(X^4)=12$: the $10+1$ classes
of $dP\times P^1_z$ plus the elliptic fibre class of F-theory 
($\rho_{K3_{12-8}}=2$).
Next the number of complex deformations of 
$X^4=dP\times_{P^1_y}{\cal B}$ can be counted
as the sum of the number of complex deformations of dP and ${\cal B}$
plus 3 (because you can use only one times the reparametrization freedom
of the common base; you can compare that procedure with the count 
in the similar case of the $CY^{(19,19)}$, where you can count $19=8+8+3$). 
Now the deformations of ${\cal B}$ 
(which is here not the $h^{(2,1)}$ as we are not on a CY) are counted as 
$2\cdot 3\cdot 10-(3+3+8)-1=45$ giving $h^{3,1}(X^4)=45+8+3=56$
and with $h^{1,1}+h^{3,1}-h^{2,1}=\frac{\chi}{6}-8=40$ of \cite{SVW} also 
$h^{2,1}=h^{3,1}-28=28$.\footnote{
One can check explicitely that $h^{2,1}=h^{2,1}({\cal B})$.
One can compute directly from the given degrees that
$e_{{\cal B}}=-48$ resp. $e_{{\cal B}^{\prime}}=-144$,
which matches with the visualization of ${\cal B}^{\prime}$ as branched 
covering of ${\cal B}$
(induced from a two-fold covering with two branch points of the base $P^1_y$),
namely $-144=2(-48)-2(24)$, where one sees that the two fixed fibers over
the two fixed points in the base are now K3's.
Now $e_{{\cal B}}
=-48$ gives with $h^{(1,0)}({\cal B})=h^{(2,0)}({\cal B})=
h^{(3,0)}({\cal B})=0$ (cf. \cite{DGW}) 
and $-48=2+2(h^{(1,1)}-h^{(2,1)})$ that $h^{2,1}({\cal B})=28=h^{2,1}$, 
which is also a number of quite visible origin:
${\cal B}$ can be considered (cf.\cite{DGW}) as the
blow-up of 
${\scriptsize \left[\begin{array}{c}P^1\\P^2\end{array}\right]}$ 
at the base
locus 
$\Gamma:={\scriptsize 
\left[\begin{array}{c|cc}P^1&2&2\\P^2&3&3\end{array}\right]}$
of the pencil of K3 surfaces. The cohomology class of $\Gamma$ is
$-2J_1-3J_2$, denoting by $J_i$ the respective classes coming from the 
factors, so $e_{\Gamma}=-\Gamma^2=-54$ (showing again 
$e_{{\cal B}}=6-54=-48$), i.e. $\Gamma$ has genus 28.}

To gain further confidence in $h^{(2,1)}
({\cal B}^{\prime})=75$,
where ${\cal B}^{\prime}=
{\tiny \left[\begin{array}{c|c}P^1&2\\P^1&2\\P^2&3\end{array}\right]}$, 
note that
not only the Euler number can be independently computed from the degrees 
to be -144, leading to the 75, but
that (cf. \cite{B}) one can follow the precise occurence of that CY through a 
conifold transition from (remarkably enough again our friend) $CY^{(19,19)}$.
To see this observe that in contrast to the easier case (considered in A.2) 
of transition to the CY 
${\scriptsize 
\left[\begin{array}{c|c}P^2&3\\P^2&3\end{array}\right]}$ in our case here
one is choosing in one del Pezzo only 8 sections leading
to the contraction of $8\cdot 9=72$ $P^1$'s 
and then detuning of $75-19=56=7\cdot 8=(8-1)\cdot (9-1)$ parameters; i.e.
the Euler number -144 is reached (from the Euler number zero of the 
$CY^{(19,19)}$) in the usual two steps: first the contraction of the 72 $P^1$'s
gives $\chi=-72$ and then the resmoothing via introduction of the three-spheres
lets it go to -144. (To implement the special features
of this transition it is useful, after replacing one elliptic fibre of the 
$CY^{19,19}$ by 
${\scriptsize \left[\begin{array}{c|c}P^1&2\\P^1&2\end{array}\right]}$, 
to consider also then the 
${\tiny 
\left[\begin{array}{c|cc}P^1&2&0\\P^1&2&0\\P^1&1&1\\P^2&0&3\end{array}\right]}$, where now 
${\scriptsize \left[\begin{array}{c|cc}P^1&2&2\\P^1&2&2\end{array}\right]}$ 
consists of 8 points instead of the 9 of 
$\left[\begin{array}{c|cc}P^2&3&3\end{array}\right]$.)\\

\setcounter{equation}{0}


\end{document}